%% file: main.tex
\documentclass[conference]{IEEEtran}
\IEEEoverridecommandlockouts
\usepackage{cite}
\usepackage{amsmath,amssymb,amsfonts}
\usepackage{algorithmic}
\usepackage{graphicx}
\usepackage{textcomp}
\usepackage{xcolor}
\usepackage{multirow}
\usepackage{nicefrac}

\newcommand{\fixme}[2]{\ifx&#2&{\color{red}#1}\else{\color{red}FIXME\{}#1{\color{red}\}}\footnote{{\color{red}#2}}\PackageWarning{Fixme}{#1: #2}\fi}

\def\BibTeX{{\rm B\kern-.05em{\sc i\kern-.025em b}\kern-.08em
    T\kern-.1667em\lower.7ex\hbox{E}\kern-.125emX}}
    
\begin{document}

\title{\textsc{OptComNet}: Optimized Neural Networks for Low-Complexity Channel Estimation}

\author{\IEEEauthorblockN{Michel van Lier\IEEEauthorrefmark{1}, Alexios Balatsoukas-Stimming\IEEEauthorrefmark{1}, Henk Corporaal\IEEEauthorrefmark{1}, and Zoran Zivkovic\IEEEauthorrefmark{2}}
\IEEEauthorblockA{\IEEEauthorrefmark{1}Eindhoven University of Technology, Eindhoven, Netherlands\\
\IEEEauthorrefmark{2}Intel Corporation, Eindhoven, Netherlands\\
E-mail: m.l.v.lier@student.tue.nl, \{a.k.balatsoukas.stimming, h.corporaal\}@tue.nl, zoran.zivkovic@intel.com
}
}

\maketitle

\begin{abstract}\label{ch:abstract}
\input{abstract.tex}
\end{abstract}


\section{Introduction}\label{ch:intro}
\input{intro.tex}

\section{Channel Estimation and Denoising}\label{ch:background}
\input{background.tex}

\section{Optimized Channel Denoising Neural~Networks}\label{ch:method}
\input{method.tex}

\section{Results}\label{ch:results}
\input{results.tex}

\section{Discussion}\label{ch:discussion}
\input{discussion.tex}

\section{Conclusions}\label{ch:conclusions}
\input{conclusions.tex}

\bibliographystyle{IEEEtran}
\bibliography{bibliography/ref}

\end{document}

%% file: abstract.tex
The use of machine learning methods to tackle challenging physical layer signal processing tasks has attracted significant attention. In this work, we focus on the use of neural networks (NNs) to perform pilot-assisted channel estimation in an OFDM system in order to avoid the challenging task of estimating the channel covariance matrix. In particular, we perform a systematic design-space exploration of NN configurations, quantization, and pruning in order to improve feedforward NN architectures that are typically used in the literature for the channel estimation task. We show that choosing an appropriate NN architecture is crucial to reduce the complexity of NN-assisted channel estimation methods. Moreover, we demonstrate that, similarly to other applications and domains, careful quantization and pruning can lead to significant complexity reduction with a negligible performance degradation. Finally, we show that using a solution with multiple distinct NNs trained for different signal-to-noise ratios interestingly leads to lower overall computational complexity and storage requirements, while achieving a better performance with respect to using a single NN trained for the entire SNR range.

%% file: intro.tex
The use of machine learning techniques, and in particular deep learning and neural networks (NNs), to tackle communications tasks has attracted significant interest in the past few years~\cite{DBLP:journals/corr/abs-1710-05312,Oshea2017,Guduz2019,Balatsoukas2019}. Many approaches have been proposed specifically for improving the performance of channel estimation in orthogonal frequency-division multiplexing (OFDM) systems, ranging from approaches which replace the entire receiver chain \cite{FC-DNN}, to model-based approaches which replace only parts of the receiver chain \cite{comnet, SwitchNet, ChannelNet}. More specifically, in~\cite{FC-DNN} the channel estimation and demodulation blocks of the receiver are replaced by a five-layer fully-connected deep NN (FC-DNN). Initial experiments show that the NN-based receiver performs better than traditional least squares (LS) and minimum mean square error (MMSE) estimation and detection, especially when very few pilot symbols are used. In ComNet~\cite{comnet}, on the other hand, a conventional LS channel estimation is first performed only for the pilot symbols and then refined using a one-layer NN. Subsequently, zero-forcing (ZF) followed by a two-layer fully-connected NN (FC-NN) is used for symbol detection. SwitchNet~\cite{SwitchNet} extends the approach of~\cite{comnet} by adding a second layer NN to the channel estimation NN. The first layer is trained on channels with a short delay spread, while the second layer is trained on channels with a large delay spread. In addition, a switch is trained to switch on the second layer when a large delay spread is detected. Finally, \cite{ChannelNet} proposes to use two convolutional NNs (CNNs) for channel estimation. In particular, the time-frequency response of a pilot-channel is treated as a low-resolution image and a super-resolution CNN is cascaded with a denoising CNN to estimate and interpolate the pilot-channel in both time and frequency to the data-channel. Simulation results show that this approach is comparable to the ideal MMSE estimator and performs better than approximated MMSE on realistic channel models. 

While FC-DNN~\cite{FC-DNN}, ComNet~\cite{comnet}, SwitchNet~\cite{SwitchNet}, and ChannelNet~\cite{ChannelNet} can outperform conventional methods in certain cases, they unfortunately also have significantly higher complexity that makes them impractical for resource-constrained mobile hardware platforms. 
Although many NN complexity reduction techniques have been proposed in the literature, such as quantization~\cite{DBLP:journals/corr/abs-1802-05668, DBLP:journals/corr/ZhouYGXC17, DBLP:journals/corr/abs-1807-10029, 8760178} and pruning \cite{hu2016network, DBLP:journals/corr/abs-1810-11809}, these methods are typically applied to classification problems and their impact on the performance of the specific regression task performed by NN-assisted channel estimation methods remains unknown.

\begin{figure}[t]
     \centering
     \includegraphics[width=0.925\columnwidth]{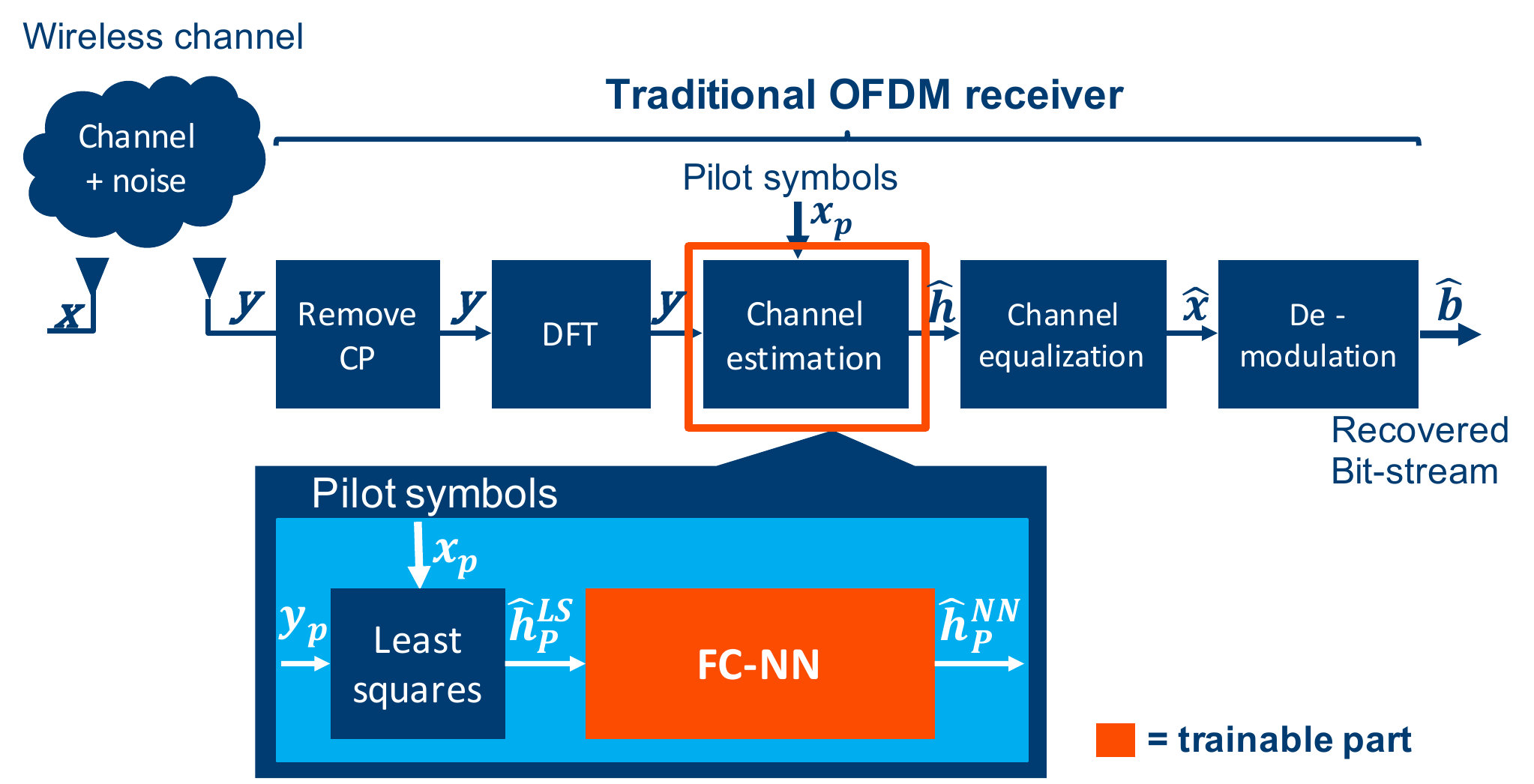}
     \caption{Example of an OFDM receiver with NN-assisted channel estimation.}
     \label{fig:ofdm_model}
	 \vspace{-0.25cm}
\end{figure}

\subsubsection*{Contributions} In this work, we examine several methods to significantly reduce the computational and memory complexity of NN-assisted channel estimators. In particular:
\begin{itemize}
	\item In Section~\ref{sec:opti_model}, we perform a systematic design-space exploration of a denoising FC-NN architecture for channel estimation and we show that a careful design of the FC-NN configuration is a necessary requirement to obtain Pareto-optimal performance-complexity trade-offs. This is a rather well-known fact in the machine learning community, but so far has received little to no attention from the communications community.
	\item In Section~\ref{sec:fp_quant}, we show that both quantization and pruning work particularly well for this application.
	\item In Section~\ref{sec:range}, our design-space exploration reveals that, interestingly, using multiple carefully designed FC-NNs trained over distinct SNR ranges outperforms a single FC-NN trained over the union of the individual SNR ranges both in terms complexity and in terms of the achieved denoising performance.
\end{itemize}


%% file: background.tex
In an OFDM system with $F$ subcarriers, such as the one shown in Fig.~\ref{fig:ofdm_model}, the transmission of a single OFDM symbol $x[f]\in\mathbb{C}$ can be modeled in the frequency domain as:
\begin{align}
	y[f]	& = h[f]x[f] + w[f], \quad f \in \{0,\hdots,F{-}1\},
\end{align}
where $y[f]\in\mathbb{C}$, $h[f]\in\mathbb{C}$, and $w[f]\in\mathbb{C}$ are the frequency-domain representations of the received signal, the transmission channel, and the additive noise on subcarrier $f$, respectively. The noise $w[f]$ is typically modeled as additive white Gaussian noise (AWGN) distributed according $\mathcal{CN}(0,\sigma ^2)$. The channels $h[f]$ at different subcarriers, however, are typically correlated. If we define $\mathbf{h} = \begin{bmatrix} h[0] & h[1] & \hdots & h[F{-}1] \end{bmatrix}$, then a common model for correlated channels is $\mathbf{h} \sim \mathcal{CN}(0,\mathbf{C}_{\mathbf{h}})$, where $\mathbf{C}_{\mathbf{h}}$ is the covariance matrix of $\mathbf{h}$.

\begin{figure}[t]
     \includegraphics[width=0.75\columnwidth]{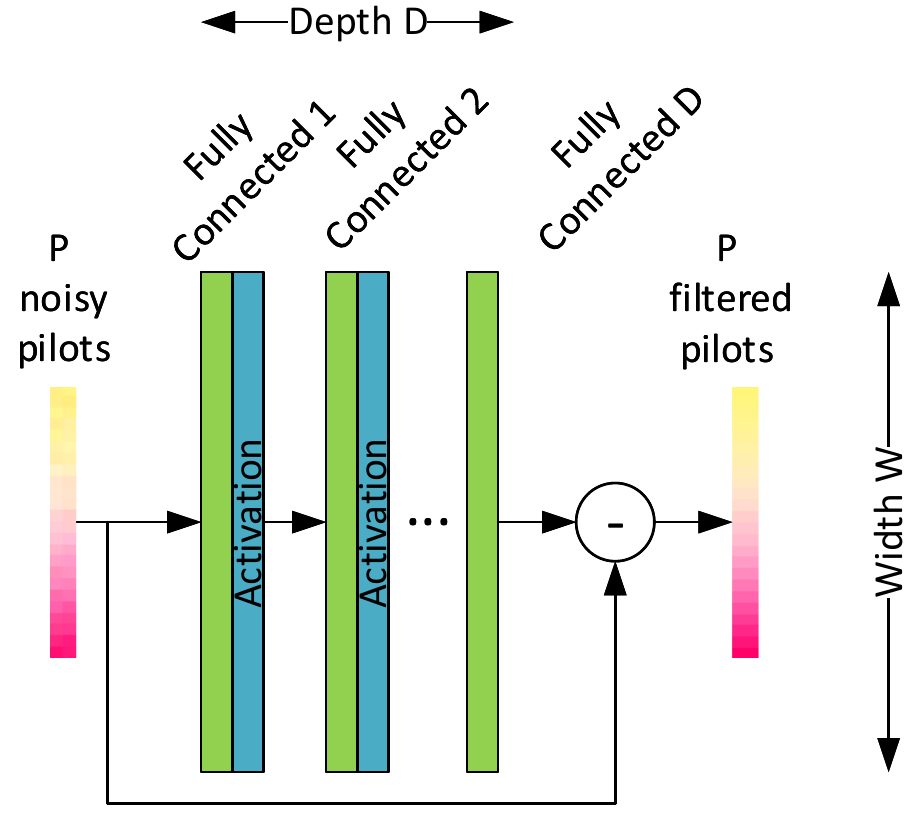}
     \centering
     \caption{Fully-connected residual NN architecture for channel estimation with $P$ pilot symbols and with a variable depth $D$ and layer width~$W$.}
     \label{fig:dnn_arch}
     \vspace{-0.25cm}
\end{figure}

\subsection{MMSE Channel Estimation}
Known pilot symbols $x[p]$, $p \in \mathcal{P} \subseteq \{0,\hdots,P{-}1\}$, are sent occasionally for the purpose of estimating the channel. LS channel estimation is given by
$\hat{\mathbf{h}}_{\mathcal{P}}^{\text{LS}}  = \mathbf{y}_{\mathcal{P}}/\mathbf{x}_{\mathcal{P}}$, 
where $\mathbf{x}_{\mathcal{P}}$ and $\mathbf{y}_{\mathcal{P}}$ denote vectors containing the elements of $x[f]$ and $y[f]$, respectively, with indices in $\mathcal{P}$ and division is performed element-wise. The LS solution is noisy in the sense that it does not take the channel correlation into account. The MMSE solution~\cite{2d_wiener}, which exploits knowledge of the covariance matrix to denoise the LS channel estimation, can be obtained as $\hat{\mathbf{h}}_{\mathcal{P}}^{\text{MMSE}} = \mathbf{C}_{\mathbf{h}_{\mathcal{P}}}(\mathbf{C}_{\mathbf{h}_{\mathcal{P}}}+\sigma^2 \mathbf{I})^{-1} \hat{\mathbf{h}}_{\mathcal{P}}^{\text{LS}}$, 
where $\mathbf{h}_{\mathcal{P}}$ denotes a vector containing the elements of $\mathbf{h}$ with indices in $\mathcal{P}$ and $\mathbf{C}_{\mathbf{h}_{\mathcal{P}}}$ denotes the covariance matrix of $\mathbf{h}_{\mathcal{P}}$.

In practice, $\mathbf{C}_{\mathbf{h}_{\mathcal{P}}}$ and $\sigma^2$ are not known and need to be estimated. Accurately estimating the covariance matrix $\mathbf{C}_{\mathbf{h}_{\mathcal{P}}}$ in particular can be challenging in practice because a large number of samples is typically required to achieve the required estimation accuracy, which is particularly restrictive in fast-fading environments and when the set $\mathcal{P}$ can change between transmissions (which is the case in, e.g., 5G NR). Furthermore, the required matrix inversion is computationally expensive.

\subsection{Channel Denoising Neural Networks}
In the previous subsection, we described MMSE as a denoising channel estimator in order to make a natural connection with NNs that are used for denoising~\cite{generic_denoise} in other areas, such as image denoising \cite{image_denoise} and speech denoising \cite{speech_denoise}. Autoencoders are a feed-forward NN often used for explaining denoising in NNs \cite{autoencoder_denoise}. They show how by disturbing input data a NN is forced to find a concise description of the data that can be used to reconstruct it. This can be seen as an explanation how denoising NNs perform the denoising function. An example of such an denoising NN is shown in Fig.~\ref{fig:dnn_arch}, where a pilot channel estimate that is distorted by noise (in our case, the LS estimate $\hat{\mathbf{h}}_{\mathcal{P}}^{\text{LS}}$) is input into the NN and a denoised, the recoverd channel estimate is produced at the output. The mean squared error (MSE) between the NN output and the true channel response $\mathbf{h}_{\mathcal{P}}$ is used to train the NN using backpropagation (BP).

\begin{table}[t]
\centering
\caption{Complexity Comparison of Existing Neural-Network-Based OFDM~Channel Estimators.}
\begin{tabular}{lll}
\hline
\textbf{Approach} & \textbf{MACs} & \textbf{Model Size} \\ 
\hline
FC-DNN \cite{FC-DNN}       					& \phantom{0}6.96 M  	& 27.85 MB \\
ComNet-BiLSTM \cite{comnet}                 & 10.40 M				& \phantom{0}2.40 MB   \\
ComNet-FC \cite{comnet}                     & \phantom{0}0.37 M    	& \phantom{0}1.25 MB  \\
SwitchNet \cite{SwitchNet}                  & \phantom{0}0.40 M    	& \phantom{0}1.82 MB  \\ 
ChannelNet \cite{ChannelNet}                & 13.15 M  				& 35.82 MB \\
\hline
\end{tabular}
\label{tbl:ce_cnn}
\vspace{-0.25cm}
\end{table}

Table~\ref{tbl:ce_cnn} shows the required number of multiply-and-accumulate (MAC) operations and the required memory to store all the NN parameters for the NNs described in \cite{FC-DNN, comnet, SwitchNet, ChannelNet}. As a comparison, conventional LS and MMSE estimators require $\mathcal{O}(P)$ and $\mathcal{O}(P^3)$ MAC-like computations, respectively, and $P$ typically ranges from a few tens to a few thousands. Moreover, LS channel estimation requires no storage, while MMSE channel estimation requires storage of a $P \times P$ matrix.

%% file: method.tex
In this section, we describe our design-space exploration methodology for the channel denoising NN, which consists of an evaluation of various NN configurations, quantization bit-widths, and neuron pruning percentages. The previous solutions we discussed in Section~\ref{ch:intro} consider denoising across both frequency and time~\cite{FC-DNN, comnet, SwitchNet, ChannelNet}. For simplicity and to enable a faster design-space exploration, in this work we focus only on denoising in the frequency dimension, but the methodology we follow can also be applied verbatim to more complex scenarios. Furthermore, frequency and time filtering can typically be performed separably~\cite{2d_wiener}. We use three performance metrics for the selection of the obtained NN configurations: 1) the model size in KB, 2) the required number of MAC operations to perform one denoising operation, and 3) the \emph{denoising gain} with respect to the LS solution, which is defined as:
\begin{align}
	G	& = 10\log_{10} \frac{\Vert\hat{\mathbf{h}}_{\mathcal{P}}^{\text{NN}} - \mathbf{h}_{\mathcal{P}}\Vert_2^2}{\Vert\hat{\mathbf{h}}_{\mathcal{P}}^{\text{LS}} - \mathbf{h}_{\mathcal{P}}\Vert_2^2}.
\end{align}
We note that future work should also examine the system-wide performance impact by e.g. measuring the systems bit-error rate (BER). Before explaining our methodology, we first briefly explain the considered channel estimation scenario. 

\subsubsection{Dataset} 
We generated a random multipath channel dataset for training. The number of distinct paths is chosen uniformly at random in $\{1,\hdots,30\}$, the (normalized) path delays are chosen uniformly at random in $[0,1]$, and the path gains are chosen uniformly at random from $0$~dB to ${-}50$~dB. The (non-normalized) delay spread for our randomly generated channels is chosen uniformly at random from $10$~ns to $1000$~ns similarly to the $6$~GHz channels provided by the 3GPP~\cite{3gpp}. The test set contains another set of similarly generated random channels that is distinct from the training set and it is also expanded with the aforementioned 3GPP channels to demonstrate the robustness and generalization power of the obtained improved denoising NNs. The pilot symbols have unit power so that the SNR is defined as $\text{SNR} = \nicefrac{1}{\sigma^2}$.

\subsubsection{Training}
The input to the NN is the LS-estimated (noisy) pilot channel  vector $\hat{\mathbf{h}}_{\mathcal{P}}^{\text{LS}}$ and the target output value is the actual channel frequency response vector $\mathbf{h}_{\mathcal{P}}$. The total loss function of the network is a weighted mean squared error (MSE) between the NN output and the channel frequency response for the $P$ pilot carriers which is defined as:
\begin{align}
	\text{MSE} & = \frac{1}{N}\sum_{n=1}^{N} ||(f(\hat{\mathbf{h}}_{\mathcal{P}}^{\text{LS}};\mathbf{\Psi})-\mathbf{h}_{\mathcal{P}})\times \text{SNR}(n)||_{2}^{2},
\end{align}
where $f$ denotes the function that the trained NN implements, $\mathbf{\Psi}$ are all the trainable parameters of the NN, $N$ is the number of training samples, $\text{SNR}(n) = \nicefrac{1}{\sigma^2(n)}$, and $\sigma^2(n)$ is the noise variance used to generate the $n$-th training sample. The scaling factor $\text{SNR}(n)$ is used to achieve a robust performance over a large SNR range with a single NN.

\subsection{NN Configuration Improvement}\label{sec:opti_model}
We start with a generic denosing FC-NN architecture shown in Fig.~\ref{fig:dnn_arch}. Since there are no clear-cut NN design guidelines in the related literature and in order to minimize the computational complexity and memory size, we explore different FC-NN configurations of this base architecture shown in Fig.~\ref{fig:dnn_arch}. Each such NN configuration is defined by the FC-NN depth $D$, the layer width $W$, and the type of activation function $ACT$. One parameter is changed at a time with a predefined step size and each floating-point model configuration is then trained from scratch. We note that, in all cases, the weights are initialized using the Glorot uniform initializer~\cite{pmlr-v9-glorot10a}. 

\subsection{Improving Across a Wide SNR range}\label{sec:range}
The NN parameter values for channel estimation generally depend on the SNR~\cite{ChannelNet} and supporting a large range of SNRs is an important requirement in wireless communications systems. Training a NN on a dataset that includes a large SNR range results in a robust performance over the entire range. However, in this case the NN needs to learn a rather complex function which requires a relatively large and high-complexity NN. A model covering a smaller range only needs to learn a simpler function and it can potentially achieve higher denoising performance for the training SNR while at the same time being smaller and lower-complexity since only a single small NN will be active at any given time. The downside of this approach is that the NN will only perform robustly close to the training SNR and multiple NNs and their parameters have to be stored if the desired SNR range is large. Thus, we use the following two approaches in our design-space exploration.

\subsubsection{\emph{\textbf{Wide SNR range}}} A single NN is trained and evaluated on a dataset that includes randomly sampled SNRs in a desired range $\left[\text{SNR}_{\min},\text{SNR}_{\max}\right]$.

\subsubsection{\emph{\textbf{Split SNR ranges}}} \label{sec:K_models} A set of $K$ models is trained, each on a single SNR. Specifically, model $k \in \{0,\hdots,K{-1}\}$, is used in the range $\left[\text{SNR}_{\min}{+}k\text{SNR}_{\text{step}},\text{SNR}_{\min}{+}(k+1)\text{SNR}_{\text{step}}\right)$, where $\text{SNR}_{\text{step}}\triangleq\frac{1}{K}\left(\text{SNR}_{\max}{-}\text{SNR}_{\min}\right)$. The training SNR is chosen as the midpoint of each range. Finally, all NN parameters are stored and only the appropriate NN is selected and used for inference depending on the SNR.

\subsection{Fixed-point Quantization}\label{sec:fp_quant}
To further reduce the memory and computation requirements, we use the  pre-trained set of floating-point Pareto-optimal NN architectures obtained from the first step described in Section~\ref{sec:opti_model} and re-train them with the addition of quantization. We quantize the weights, the biases, and the activations in each layer of the NN. Recent works have shown that deterministic quantization outperforms stochastic quantization~\cite{DBLP:journals/corr/abs-1806-08342}. It has been shown that FC-NNs and CNNs generally do not need a large bit-width to maintain good performance, but that the limited dynamic range of fixed-point numbers may be problematic. From our dynamic range analysis, we found  that  we  could  get  the  same  precision  with  fewer  bits  using  an  asymmetric  quantization scheme. Therefore, to maximize the performance we use deterministic uniform affine quantization to quantize a set of floating-point variables $\mathcal{X}$ to a set of fixed-point variables $\mathcal{X}_{Q}$ with a bit-width $Q$. Each element $x \in \mathcal{X}$ is quantized as:
\begin{align}
    x_{\text{int}}  & = \text{round}\left(\dfrac{x}{\Delta}\right), \\
    x_{Q}       	& = \text{clamp}(x_{\text{int}},0,2^{Q}-1),
\end{align}
where $ \text{clamp}(x,a,b) \triangleq \max(\min(x,b),a)$ with $a \leq b$ and the scaling factor $\Delta$ is defined as:
\begin{equation}
    \Delta = \frac{\max(\mathcal{X})-\min(\mathcal{X})}{2^{Q}-1}.
\end{equation}
The derivative of the quantization function is zero almost everywhere, so we use a straight-through estimator for BP~\cite{NIPS2015_5647}.

The weights in different layers can have significantly different distributions. For the same bit-width $Q$, smaller parameter ranges can be represented with higher precision than larger parameter ranges. Due to the different distributions for the parameters in each layer, using a common bit-width $Q$ and range $\Delta$ for all layers and parameter types (i.e., weights, biases, and activations) is generally not ideal. By allowing distinct ranges $\Delta$ and bit-widths $Q$ for each layer and parameter type, we can reduce the quantization error, at a small cost of storing three scaling values per layer. We note that this overhead is included in the model size results and we report the worst-case bit-width in our results in Section~\ref{ch:results}.

\subsection{Neuron Pruning}
To reduce both memory and computational complexity even further, we also apply a pruning approach based on the average percentage of zeros (APoZ) proposed in~\cite{hu2016network}, as a showcase. The APoZ is defined as the percentage (or, equivalently, fraction) of zero-valued neuron activations of a particular neuron over a set of $N$ inputs. Let $O^{(l)}_{c}(n)$ denote the output of neuron $c$ in layer $l$ for input sample $n$. Then, the $\text{APoZ}^{(l)}_{c}$ of neuron $c$ in layer $l$ is defined as:
\begin{align}
    \text{APoZ}^{(l)}_{c} & = \frac{1}{N}\sum_{n=1}^{N} \mathbb{I}\left(O^{(l)}_{c}(n) = 0\right),
\end{align}
where $\mathbb{I}(\cdot)$ is an indicator function. All neurons in the NN for which $\text{APoZ}^{(l)}_{c}$ is smaller than some threshold $t$ are pruned. The threshold $t$ is gradually increased in steps of $1$\% and pruning is followed by retraining to compensate for any performance degradation. To keep the design space size reasonable, only the set of pre-trained Pareto-optimal quantized NNs obtained from the procedure of Section~\ref{sec:fp_quant} are considered and re-trained with pruning.

%% file: results.tex
\begin{figure}[t]
     \centering
     \includegraphics[width=0.95\columnwidth]{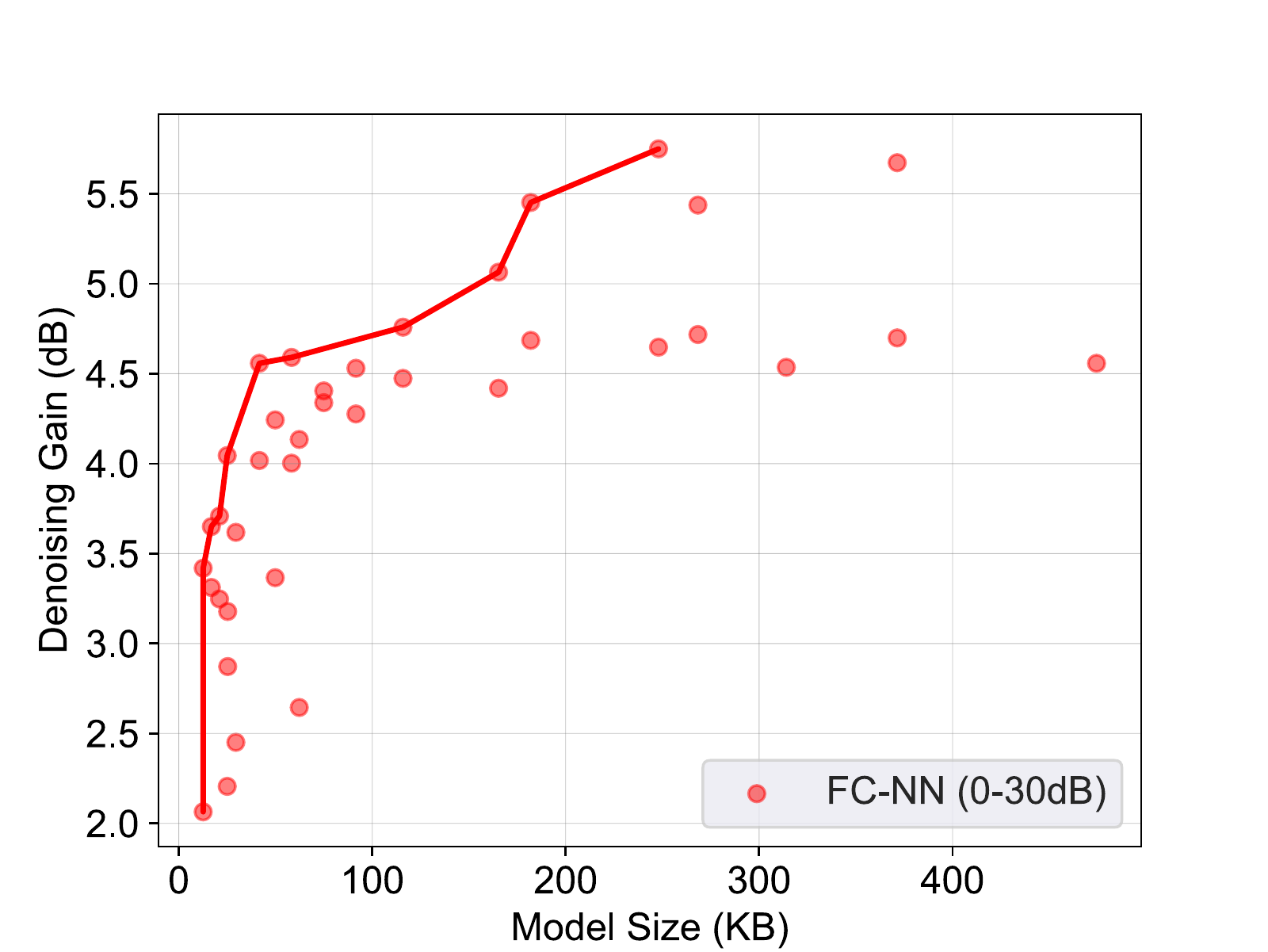}
     \caption{Design-space exploration where each point represents a single FC-NN that is used for the entire SNR range from 0 dB to 30 dB. The line represents the Pareto frontier.}
     \label{fig:quant_models_1}
     \vspace{-0.25cm}
\end{figure}
In this section, we present and interpret the results of our design-space exploration. We note that our goal is not to directly compare our results to previous work but rather to compare the various complexity-reduction methods in a well-defined and fully-controlled environment against our own baseline solution. Moreover, the works of~\cite{FC-DNN, comnet, SwitchNet, ChannelNet} consider a different setting than our work and the methodology that we explore can also be applied verbatim to these works.

We consider SNRs ranging from $0$~dB to $30$~dB, which are practically relevant for a modern OFDM-based system that uses adaptive modulation and coding. Each OFDM symbol contains $72$ data symbols and $24$ pilot symbols. 
A preliminary exploration showed that, for $W>160$ and $D>6$ no denoising performance increase is obtained, while for $W<32$ and $D<2$ the performance is far from satisfactory. 
Moreover, only multiples of $32$ are explored for $W$ for two reasons:  1) to reduce the design space, as time-consuming training is required in order to obtain the denoising performance of each considered NN architecture and 2) because communications hardware typically uses vector processors where the vector sizes are powers two. Finally, we consider the two activation functions which are used in the related works and are also generally the most widely used, namely tanh and ReLU. 
As such, the explored parameter ranges are:
\begin{align}
    W & \in \{32,64,96,128,160\}, \\
	D & \in \{2,4,5,6\}, \\
	\text{ACT} & \in \{\text{tanh}, \text{ReLU}\}.
\end{align}
For the quantization bit-width, we consider $Q \in \{8,10,12,16,32\}$ in order to cover commonly supported bit-widths and some additional in-between values. 

\begin{figure}[t]
     \centering
     \includegraphics[width=0.95\columnwidth]{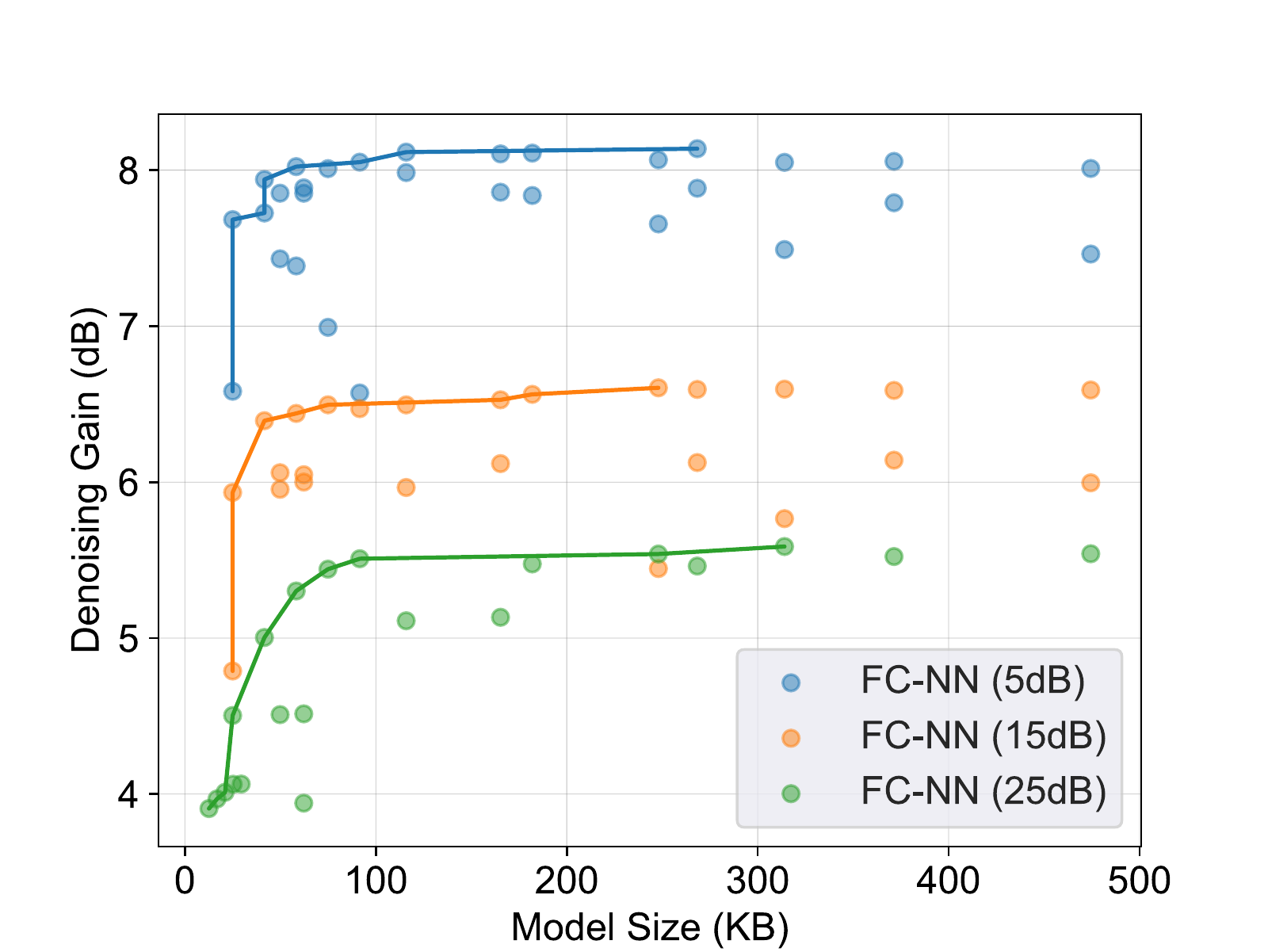}
     \caption{Design-space exploration where each point represents three distinct FC-NNs that are used for the SNR range from 0 dB to 30 dB. The lines represent the Pareto frontiers for each FC-NN.}
     \label{fig:quant_models_3}
    \vspace{-0.25cm}
\end{figure}
\subsection{NN Configuration Improvement}
Fig.~\ref{fig:quant_models_1} and Fig.~\ref{fig:quant_models_3} show the results of our design-space exploration only for the floating-point FC-NN configuration (i.e., not considering quantization or pruning yet) for the cases where a single FC-NN and $K=3$ distinct FC-NNs are used for the SNR range of interest, respectively. We observe that the model size versus performance trade-off curve is steep in both cases and levels off quickly after some model size. This shows that it is crucial to carefully select an appropriate NN configuration in order to maximize the denoising gain for a given model size constraint or to minimize the model size for a given denoising gain target.

\begin{table*}[t]
\begin{center}
\caption{Comparison of the average denoising performance, MACs, and model size over the SNR range from $0$ dB to $30$ dB for various numbers $K$ of distinct NNs.}
\label{tab:split_snr_comp}
\setlength{\tabcolsep}{1.25em}
\begin{tabular}{c|lll|lll|lll}
			& \multicolumn{3}{c|}{$Q=32$}     & \multicolumn{3}{c|}{$Q=10$}      & \multicolumn{3}{c}{$Q=10$ \& pruning} \\
Models		& Perf.	& MACs & Model Size & Perf.	& MACs & Model Size & Perf.	& MACs & Model Size \\
\hline
1         	& 5.75 dB & 62000 & 248.00 KB 			& 5.40 dB & 62000 & 77.50 KB & 5.12 dB & 33159 & 41.44 KB \\
2           & 6.01 dB & 10416 & \phantom{0}83.00 KB  	& 5.80 dB & 10416 & 26.00 KB & 5.79 dB & \phantom{0}8547 & 21.40 KB \\
3           & 6.60 dB & 10416 & 125.00 KB  			& 6.58 dB & 10416 & 39.00 KB & \textbf{6.47 dB} & \phantom{0}\textbf{8547} & \textbf{32.00 KB} \\
4           & \textbf{6.83 dB} & \textbf{10416} & \textbf{166.70 KB} 			& \textbf{6.82 dB} & \textbf{10416} & \textbf{52.10 KB} & 6.74 dB & \phantom{0}8547 & 42.70 KB
\end{tabular}
\end{center}
\scriptsize $^1$For reference, the average denoising gain for the ideal MMSE estimator with perfect knowledge of the channel covariance matrix in this SNR range is $9.1$ dB and the denoising gain of a more realistic MMSE estimator that uses $100$ samples to estimate the channel covariance matrix is $5.22$ dB.
\vspace{-.25cm}
\end{table*}

\subsection{Improving Across a Wide SNR range}\label{ch:results_float}
By comparing Fig.~\ref{fig:quant_models_1} and Fig.~\ref{fig:quant_models_3}, we also observe that, as expected, the $K=3$ distinct FC-NNs have a better denoising gain than the single FC-NN over the SNR ranges where they are used (Section~\ref{sec:K_models}). Moreover, we observe that a lower denoising gain is achieved at low SNRs than at high SNRs due to the limiting effect of the additive Gaussian noise at low SNRs. The average performance of the single FC-NN is limited by the performance at low SNRs. 

The results for different values of $K$ (with $K=1$ corresponding to the single large SNR range FC-NN) are summarized in the second column of Table~\ref{tab:split_snr_comp}, where we show the average denoising performance over the entire range of interest. Moreover, for $K>1$ we report the worst-case number of MACs for the largest FC-NN and the sum of the model sizes of all $K$ FC-NNs. We observe that, without quantization or pruning, using $K=4$ results in a $1.08$~dB better denoising gain with $83$\% fewer MACs and a $33$\% smaller model size compared to the $K=1$ case.

\begin{figure}[t]
     \centering
     \includegraphics[width=0.95\columnwidth]{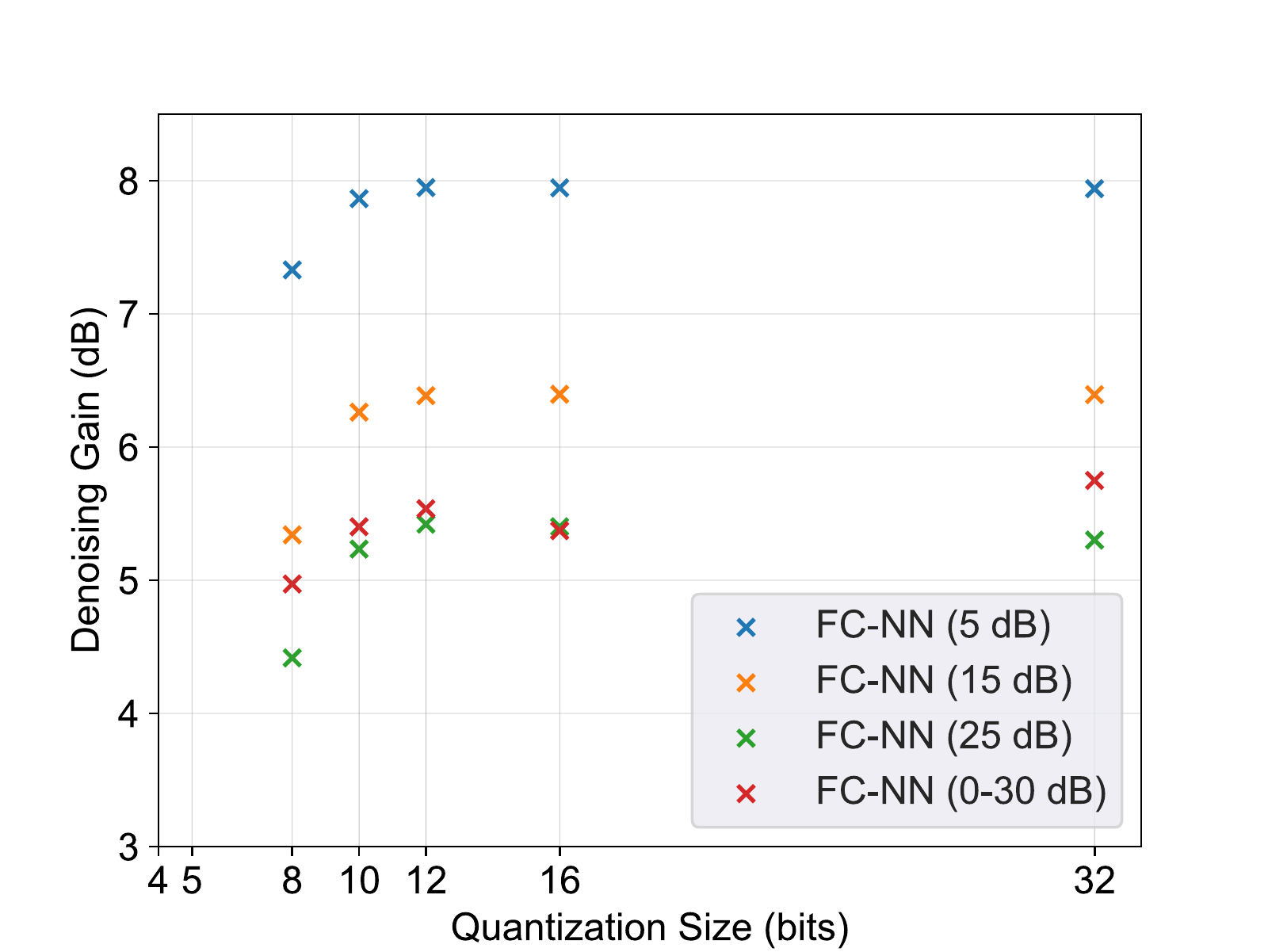}
     \caption{Performance of Pareto-optimal FC-NNs after quantization for different bit-widths $Q \in \{8,10,12,16,32\}$.}
     \label{fig:quant_robust}
     \vspace{-0.25cm}
\end{figure}

\subsection{Fixed-Point Quantization}\label{ch:results_quant}
In Fig.~\ref{fig:quant_robust} we show the denoising gain as a function of the quantization bit-width for the cases where a single FC-NN and $K=3$ distinct FC-NNs are used for the SNR range of interest. For this evaluation, we selected Pareto-optimal FC-NN configurations for each case from the previous design-space exploration step from Section~\ref{ch:results_float}. We observe that all considered FC-NNs are quite robust to quantization down to a bit-width of $Q=10$ bits, while a significant performance degradation starts appearing for $Q = 8$ bits. Interestingly, all FC-NNs have a very similar robustness with respect to quantization, although the single FC-NN seems to suffer a slightly larger loss when going from $Q=32$ to $Q=10$.

In the third column of Table~\ref{tab:split_snr_comp}, we observe that, with fixed-point quantization using $Q=10$, selecting $K=4$ results in a $1.42$~dB better denoising gain with $83$\% fewer MACs and a $33$\% smaller model size compared to the $K=1$ case.

\subsection{Neuron Pruning}
\begin{figure}[t]
     \centering
     \includegraphics[width=0.95\columnwidth]{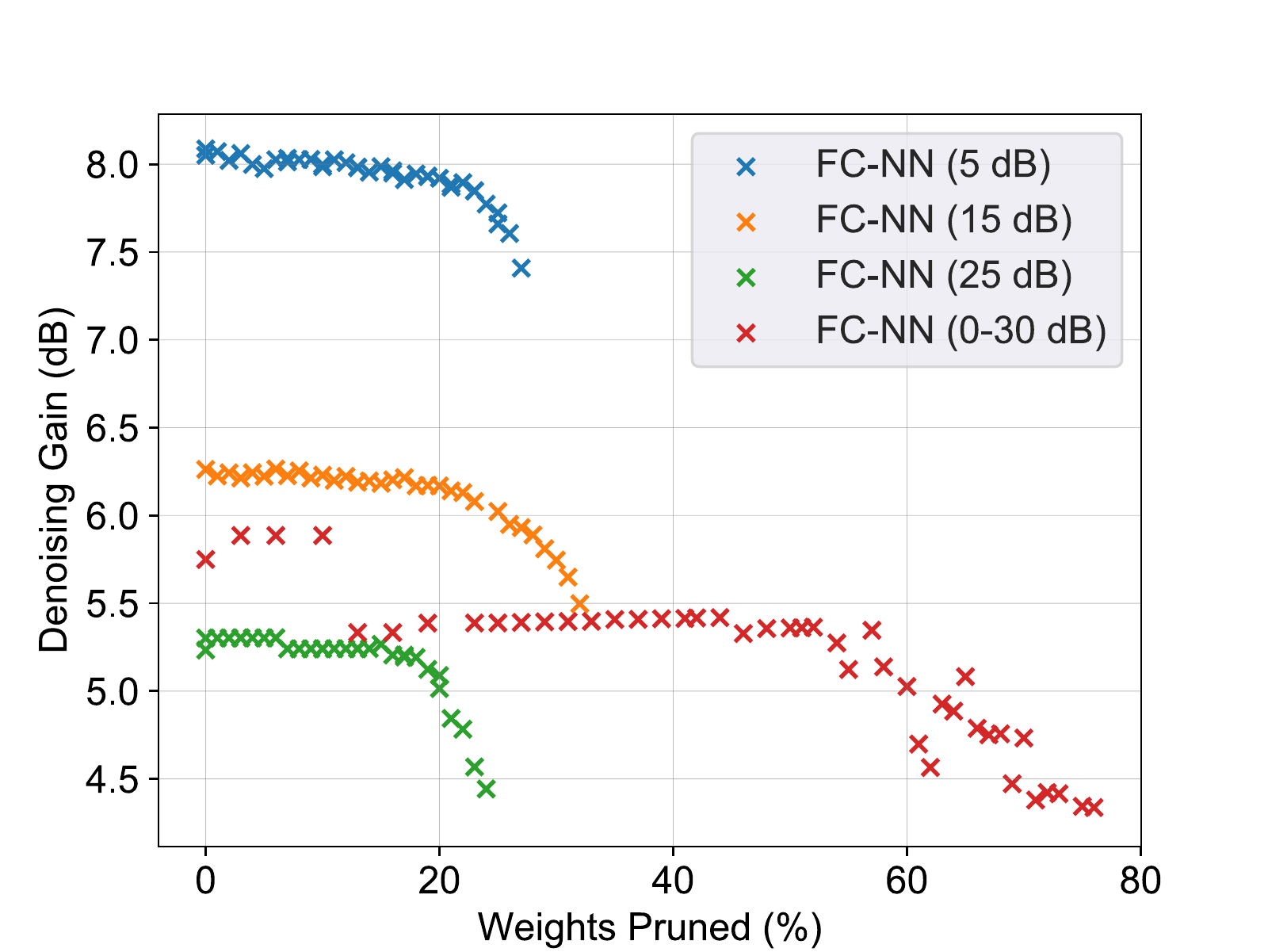}
     \caption{Performance of Pareto-optimal FC-NNs after neuron pruning for different thresholds $t$ resulting in different percentages of pruned neurons.}
     \label{fig:pruning_robust}
     \vspace{-0.25cm}
\end{figure}

In Fig.~\ref{fig:pruning_robust} we show the denoising gain as a function of the percentage of pruned neurons for the cases where a single FC-NN and $K=3$ distinct FC-NNs are used for the SNR range of interest. Different percentages of pruned neurons are obtained by varying the pruning threshold $t$ and pruning is applied to the Pareto-optimal quantized FC-NNs from Section~\ref{ch:results_quant}. We observe that the single FC-NN is initially sensitive to pruning and shows a performance loss of approximately $0.5$ dB when pruning $10$\% of the neurons. However, after that point the single FC-NN is very robust to pruning and approximately $48$\% of the neurons can be pruned without a significant additional performance degradation. For the case where $K=3$, all FC-NNs degrade much more gracefully with increasing neuron pruning percentages, but only about $20$\% pruning is possible without a significant performance degradation.

In the fourth column of Table~\ref{tab:split_snr_comp}, we observe that, when combining both quantization and pruning, selecting $K=3$ results in a $1.35$~dB better denoising gain with $74$\% fewer MACs and a $23$\% smaller model size compared to the $K=1$ case. We note that, based on the previous results, $48$\% pruning is used for $K=1$ and $20$\% pruning is used for $K \geq 1$. Due to the more limited pruning that is possible for the $K>1$ cases, in this scenario the solution with $K=4$ actually has a $3$\% larger model size than the solution with $K=1$, but it can achieve a $1.62$~dB better denoising gain.

%% file: discussion.tex
In this section, we briefly discuss some other common NN techniques that we considered in our design-space exploration but that did not give significant improvements. The LS channel estimates are complex-valued, meaning that we need to split the complex input into real and imaginary parts in order to use standard real-valued NNs. We attempted to directly use complex-valued NNs, but did not obtain any noteworthy improvements in neither performance nor complexity. Moreover, to see if we could reduce the model size of the $K>1$ models we tried to re-use some of the first layers across different SNR ranges, as we hypothesized that the features in the first layers might be similar. This did not improve the performance and, in some cases, even made it worse. Finally, we also attempted to use one-dimensional CNNs, but this resulted in worse denoising performance with higher complexity.

%% file: conclusions.tex
In this work, we performed a systematic design-space exploration of denoising NNs for channel estimation in OFDM systems. We showed that carefully exploring and selecting the NN configuration is necessary to obtain Pareto-optimal performance-complexity trade-offs. Moreover, we confirmed that quantization and pruning are effective complexity-reduction methods for this application. Specifically, when applied to Pareto-optimal model configuration, a bit-width of $Q=10$ bits is sufficient for the examined channel denoising scenario. After selecting and quantizing a model often almost $50$\% of the neurons can be pruned with negligible performance degradation in comparison to the floating-point Pareto optimal counterpart. Finally, we showed that, during all steps of our design-space exploration using multiple distinct NNs trained for different SNR ranges can be more effective than using a single NN trained over the entire SNR range of interest. For example, when using both quantization and pruning, $K=3$ distinct NNs have a $1.35$~dB better average denoising gain while also being $74$\% less computationally complex and $23$\% smaller in terms of the model size.